\documentclass[prl,twocolumn,superscriptaddress]{revtex4}
\usepackage{graphicx}
\usepackage{dcolumn}
\usepackage{bm}

\begin{document}

\title{Auxiliary Field Diffusion Monte Carlo calculation of nuclei with $A\le 40$ with tensor interactions.}

\author{S. Gandolfi}
\email{gandolfi@science.unitn.it}
\affiliation{Dipartimento di Fisica, Universit\'{a} di Trento, 
via Sommarive 14, I--38050 Povo, Trento Italy}
\affiliation{INFN, Gruppo collegato di Trento, Universit\'{a} di Trento, 
via Sommarive 14, I--38050 Povo, Trento Italy}
\author{F. Pederiva}
\affiliation{Dipartimento di Fisica, Universit\'{a} di Trento, 
via Sommarive 14, I--38050 Povo, Trento Italy}
\affiliation{INFM {\sl DEMOCRITOS} National Simulation Center, Via Beirut 2/4
I-34014 Trieste,  Italy}
\author{S. Fantoni}
\affiliation{International School for Advanced Studies, SISSA
Via Beirut 2/4 I-34014 Trieste, Italy}
\affiliation{INFM {\sl DEMOCRITOS} National Simulation Center, Via Beirut 2/4
I-34014 Trieste,  Italy}
\author{K. E. Schmidt}
\affiliation{Department of Physics, Arizona State University, Tempe, AZ 85287 USA}

\date{\today}

\begin{abstract}
We calculate the ground-state energy of $^4$He, $^8$He, $^{16}$O, and
$^{40}$Ca using
the auxiliary field diffusion Monte Carlo method in the fixed phase 
approximation and the Argonne $v_6'$
interaction which includes a tensor force.
Comparison of our light nuclei results to
those of
Green's function Monte
Carlo calculations shows the accuracy of our method for both open and
closed shell nuclei. We also apply it
to $^{16}$O and $^{40}$Ca to show that quantum Monte Carlo methods are now
applicable to larger nuclei.

\end{abstract}


\maketitle

The nuclear many--body problem has not found a satisfactory solution yet.
As opposed to what happens for other systems, one has to face both
the absence of a definitive scheme for the description of nucleon--nucleon
forces, and with the extreme complexity induced by the fact that
forces are state dependent. Recently there have been  a few
attempts to reduce the problem to a more fundamental level by exploiting
in full the scheme of  Effective Field Theory (EFT). However, this approach,
at present, is applicable only to very small nuclei\cite{borasoy07}. Integration
of subnuclear degrees of freedom in EFT now provide interactions
which are almost as accurate in describing scattering data as 
the most popular realistic interactions\cite{entem03}. However, the assessment of the 
quality of different potentials in many--body sistems remains problematic.
Quantum Monte Carlo 
is the only available tool that provides estimates of physical observables 
with an accuracy
comparable to that obtained by means of few--body techniques
\cite{kamada01}, but in a wider range of nuclear masses. 
QMC calculations might therefore help to discuss and gauge 
the validity of the proposed interactions without the bias implied
by the use of approximate methods. However, the operatorial
structure of such potentials, both phenomenological and EFT,
prevented to push calculations beyond $A=12$\cite{pieper05}.
QMC methods deal with the exponential increase in the
computational time with particle number by sampling the degrees of freedom
to evaluate sums and integrals. The variational and
Green's function Monte Carlo methods\cite{pieper05}
are some of
the most successful methods for calculating the properties
of light nuclei. There the
spatial degrees of freedom are sampled, but the spin isospin degrees
of freedom of the nucleons are explicitly summed and not sampled. The exponential
growth of the spatial degrees of freedom is controlled, but since
there are four spin-isospin states per nucleon, the
computations grow exponentially -- roughly as four raised to
the number of nucleons. To make the method computationally efficient,
the spin-isospin degrees of freedom must also be sampled.
That is what the Auxiliary Field Diffusion Monte Carlo (AFDMC)\cite{schmidt99}
does in the most efficient way known today.

Here we will demonstrate that AFDMC can be used to solve for the energy of
nuclei with nucleons interacting via
a somewhat
simplified two-body interaction
which, however, does contain the tensor interaction that comes from the one-pion
exchange potential, and already includes all the terms which make
standard approaches unpractical. 
This interaction term is indeed the most demanding one, together with the short 
range repulsion, in solving the nuclear many-body Schr\"odinger equation. Its inclusion 
in the Hamiltonian provides one of the most severe tests of the efficiency of a quantum 
Monte Carlo algorithm. Adding the neglected spin-orbit terms and the three-body potential 
is not expected to change the main conclusion of this paper, namely that
AFDMC is applicable with the same accuracy to both nuclei and nuclear matter.

The AFDMC method which includes
a path constraint to control the fermion
sign problem, has given good results for pure neutron
matter\cite{fantoni01,sarsa03}, for 
neutron-drops\cite{pederiva04} and for the valence neutrons of
neutron-rich nuclei\cite{gandolfi06}. 
For nuclei, the strong tensor force in the isosinglet channel makes sampling
the spin-isospin states more difficult, 
leading to unsatisfactory results when np and pp interactions are active.
In this letter we demonstrate that AFDMC 
in the fixed phase approximation overcomes this problem, and, as a consequence,
can be applied to calculate binding
energy of large nuclei and nuclear matter with realistic interactions.
Most of our results are with the Argonne $v_6'$
interaction\cite{wiringa02},
a simplified version of the Argonne
$v_{18}$ potential \cite{wiringa95} truncated to the first six operators 
and modified to describe the binding energy of deuteron.
We calculate the energy of the alpha particle and the open shell nucleus
$^8$He and test the accuracy of
our results 
by comparing them to those from GFMC\cite{wiringa02}.
We apply the same algorithm to the study of 
binding energy of $^{16}$O first with the Argonne $v_{14}$ interaction
truncated to include only six operators
to compare our results with those of
Cluster Variational Monte Carlo\cite{pieper92} and 
Fermi Hypernetted Chain in the Single Operator Chain approximation
(FHNC/SOC). Finally, we calculate,
the ground-state energy of the
closed-shell nuclei $^{16}$O and $^{40}$Ca
using the Argonne $v_6'$ interaction.

Monte Carlo methods are most efficient when applied to sums and integrals
with positive kernels. These can be interpreted as probabilities and
probability densities. We have chosen to use the Argonne series of
potentials\cite{wiringa02} because they are substantially spatially
local with only a few derivative terms. The short-time
Green's functions that we need are then readily calculated and sampled.

Our Hamiltonian is
\begin{eqnarray}
H = \sum_i \frac{p^2_i}{2m} + \sum_{i<j}\sum_{n=1}^M v_n(r_{ij}) O^{(n)}(i,j)
\nonumber
\end{eqnarray}
where $i$ and $j$ label the two nucleons, $ r_{ij}$ is the distance
separating the two nucleons, and the $O^{(p)}$ include spin and isospin
operators, where M is the number of operators
(i.e. 18 in $v_{18}$ models).
The mass $m^{-1} = (m_p^{-1}+m_n^{-1})/2$ where $m_p$ and $m_n$ are
the proton and neutron masses.
We use the Argonne $v_6'$\cite{wiringa02} model where the two-body
potential is
reprojected from the Argonne $v_{18}$ to the $M=6$ level.
The six $O^{(n)}(i,j)$ terms are
the $1$,
$\vec \tau_i\cdot \vec \tau_j$,
$\vec  \sigma_i \cdot \vec \sigma_j$,
$(\vec \sigma_i \cdot \vec \sigma_j)(\vec \tau_i\cdot \vec \tau_j)$,
$S_{ij}$,
and
$S_{ij} \vec \tau\cdot \vec \tau_j$,
where $S_{ij}$ is the tensor
operator
$ 3 \vec \sigma_i \cdot \hat r_{ij} \vec \sigma_j \cdot
\hat r_{ij} -\vec \sigma_i \cdot \vec \sigma_j$.
The $\vec \tau_i$ and $\vec \sigma_i$
are the Pauli matrices for the isospin and spin of particle $i$.
The inclusion of neutron-proton mass difference, electromagnetic interactions,
spin-orbit interactions, and three-body potentials can be done with
an increase in complexity. However the bulk of the binding energy comes
from the $v_6$ terms which include the one-pion exchange
parts of the potential.

Traditionally, ground-state quantum Monte Carlo calculations begin by
using a variational calculation to optimize a trial wave function. This
trial wave function is then used to guide the sampling of the
random walk in diffusion or Green's function Monte Carlo. A typical form
for a good trial function has a model function $|\Phi\rangle$
given by a small linear combination of antisymmetric
products of orbitals multiplied by a symmetrized product of two-body
operator correlations. Evaluating this symmetrized product
trial function at the spatial
positions $R$ and spin-isospin values $S$ gives the expression
\begin{equation}
\langle R, S |\Psi_{SP}\rangle =
\langle R, S|
{\cal S} \prod_{i<j} \left [ \sum_{p=1}^M f^{(p)}(r_{ij})
O^{(p)}(i,j) \right ]|\Phi\rangle \,.
\end{equation}
Unfortunately, the evaluation of
this wave function requires exponentially increasing
computational time with the number of particles. Since the evaluation for
all spin-isospin values has the same computational complexity, light
nuclei variational and Green's function Monte Carlo calculations sum
the spin-isospin degrees of freedom.

Since for large numbers of particles we cannot evaluate these trial functions,
we use much simpler wave functions which contain only the central Jastrow
correlation. The evaluation of our simpler wave function require order $A^3$
operations to evaluate the Slater determinants and $A^2$ operations for
the central Jastrow. Since many important correlations are neglected in
these simplified functions, we use the Hamiltonian itself to define the
spin sampling.

Specifically our trial function is
\begin{eqnarray}
\langle R S |\Psi_T \rangle =
\left [\prod_{i<j} f^c_{ij} \right ]
{\cal A} \left [ \prod_i \phi_i(\vec r_i - R_{\rm cm} , s_i) \right ] \,.
\end{eqnarray}
where $\cal A$ is an antisymmetrization operator, $\phi$ are single particle
space and spin-isospin orbitals, built from combinations of
radial functions, spherical harmonics and spinors.
$R_{\rm cm} = A^{-1} \sum_{i=1}^A \vec r_i$
is the center of mass of the nucleus. 
The Jastrow function is
the central part of the 
FHNC/SOC\cite{pandharipande79} correlation operator 
$\hat F_{ij}$ which minimized the energy 
of nuclear matter at $\rho_0$=0.16 fm$^{-3}$.

Radial orbitals are calculated in the self-consistent potential generated by the
Hartree-Fock algorithm with 
Skyrme's effective interactions of Ref. \cite{xinhua97}
that has been used to study 
light nuclei. 
Given a set of positions and spinors, the antisymmetrization produces
a determinant of single particle orbitals. For open-shell nuclei, a sum of 
several determinants is used to build a trial wavefunction of a total 
angular momentum $J$ that describes
the nucleus.
The determinants are multiplied by the central Jastrow factor to
give the value of our trial function.


The AFDMC method works much like diffusion Monte Carlo\cite{schmidt99,
fantoni01,pederiva04,gandolfi06}. The wave
function is defined by a set of what we call walkers. Each walker is a set
of the $3A$ coordinates of the particles plus 
$A$ normalized four component spinors representing the spin-isospin state.
The imaginary time propagator for the
kinetic energy and the spin-independent part of the potential is identical
to that used in standard diffusion Monte Carlo. The new positions are sampled from
a drifted Gaussian with a weight factor for branching given by the local
energy of these components. Since these parts do not change the spin state,
the spinors will be unchanged by these parts of  propagator.

To sample the spinors we first use a Hubbard Stratonovich transformation to
write the propagator as an integral over auxiliary fields of a separated
product of single particle spin-isospin operators. We then sample the
auxiliary field value, and the resulting sample independently
changes each spinor for each particle in the sample, giving a new sampled
walker.

Specifically we write the $v_6$ interaction as
\begin{eqnarray}
\label{eq.pot}
V &=& \sum_{i<j} v_1(r_{ij}) + V_{sd}
\nonumber\\
V_{sd} &=&
\frac{1}{2}\sum_{i\alpha,j\beta} \sigma_{i\alpha}
A^{(\sigma)}_{i\alpha,j\beta} \sigma_{j\beta}
+\frac{1}{2}\sum_{i\alpha,j\beta} \sigma_{i\alpha}
A^{(\sigma\tau)}_{i\alpha,j\beta} \sigma_{j\beta}
\vec \tau_i \cdot \vec \tau_j \nonumber \\
&+&\frac{1}{2}\sum_{i,j}
A^{(\tau)}_{i,j}
\vec \tau_i \cdot \vec \tau_j.
\end{eqnarray}
Where $v_1(r)$ is the central interaction, and
$V_{sd}$ is the spin-isospin dependent part.
The $A$ matrices depend only on the positions of the particles.
They are zero when $i=j$ and they are real and symmetric
so that they have real eigenvalues $\lambda_n^{(\sigma)}$,
$\lambda_n^{(\sigma \tau)}$, $\lambda_n^{(\tau)}$ and corresponding
real normalized eigenvectors $\psi_n^{(\sigma)}(i,\alpha)$,
$\psi_n^{(\sigma\tau)}(i,\alpha)$,$\psi_n^{(\tau)}(i)$.
The spin-dependent potential can be written as a sum of squares of
single-particle operators as
\begin{eqnarray}
\label{eq.vsd}
V_{sd} &=& \frac{1}{2} \sum_{m=1}^{15A} \lambda_m O_m^2
\end{eqnarray}
where the $15A$ operators
\begin{eqnarray}
O_{n}^{(\sigma)} &=& \sum_{i\alpha} \sigma_{i\alpha}\psi^{(\sigma)}_n(i,\alpha)
\nonumber\\
O_{n\alpha}^{(\sigma\tau)} &=&
\sum_{i\beta} \tau_{i\alpha}\sigma_{i\beta} \psi^{(\sigma \tau)}_n(i,\beta)
\nonumber\\
O_{n\alpha}^{(\tau)} &=& \sum_{i} \tau_{i\alpha} \psi^{(\tau)}_n(i)
\end{eqnarray}
are labeled in a convenient order of the eigenvectors and
the corresponding eigenvalues.

We apply the Hubbard-Stratonovich transformation to write
\begin{eqnarray}
e^{-\frac{1}{2} \Delta t \lambda O^2}
= \frac{1}{\sqrt{2\pi}}\int_{-\infty}^\infty dx
e^{-\frac{x^2}{2}+\sqrt{-\lambda\Delta t} x O} \,,
\end{eqnarray}
where $x$ is an auxiliary field. Each of the 15A terms in Eq. \ref{eq.vsd}
requires an auxiliary field. We write the short time
approximation of the spin-dependent propagator as
\begin{eqnarray}
e^{-V_{sd}\Delta t} =
\int dX
\exp\left [ -\sum_{n=1}^{15A} \left (
\frac{x^2_n}{2}
- x_n \sqrt{-\lambda_n\Delta t} O_n \right ) \right ]
\end{eqnarray}
where $dX \equiv \prod_{n=1}^{15A} \frac{dx_n}{\sqrt{2\pi}}$ and
we drop commutator terms which are
higher order than $\Delta t$ on the right
hand side.

Once the Hubbard-Stratonovich variables have been sampled,
the resulting propagator acting on a walker (i.e. positions
and spinors) gives a single new walker.

Since the trial function evaluated at the walker spin-isospin and
position can be complex, we use a
fixed-phase approximation\cite{ortiz93}.
We importance sample the auxiliary field variables $x_m$ by writing
\begin{eqnarray}
&& \frac{x^2_n}{2}
+ x_n \sqrt{-\lambda_n\Delta t} O_n
\nonumber\\
&& = \frac{x^2_n}{2}
+ x_n \sqrt{-\lambda_n\Delta t} \langle O_n\rangle
+ x_n \sqrt{-\lambda_n\Delta t} (O_n-\langle O_n\rangle)
\nonumber\\
\end{eqnarray}
where
$\langle O_n\rangle =\langle \Psi_T |O_n|R,S\rangle/\langle \Psi_T|R,S\rangle$
is the mixed expectation value. The first two terms are then combined
to form a shifted contour Gaussian as in Ref. \cite{zhang03}.
Applying the fixed phase approximation, instead of the previously used
constrained path,  the walker weight can be reexpressed in terms of the local energy
\begin{equation}
E_L(R,S) = {\rm Re}
\frac{\langle \Psi_T |H |R S\rangle}{\langle \Psi_T |R S \rangle} \,.
\end{equation}
This change to the original algorithm was necessary to overcome the unphysical 
discrepancies observed in earlier AFDMC work on nuclei with tensor forces \cite{Schmidt03}.

Our algorithm becomes: i) sample $|R,S \rangle$ initial walkers  
from $|\langle \Psi_T | R, S \rangle|^2$ using Metropolis Monte Carlo;
ii) propagate in the usual DMC way with a drifted
Gaussian for a time step; iii) diagonalize, for each walker, 
the potential matrices $A^{(\sigma)}$, $A^{(\tau)}$ and
$A^{(\sigma\tau)}$; iv) sample the corresponding shifted contour auxiliary field
variables and update the spinors. The new
walker has a weight given by $\exp(-E_L(R',S')\Delta t)$.


\begin{table}
\caption{The Ground-State energies of the alpha particle and of $^8He$
calculated with different methods 
using the Argonne $v_6'$ interaction. The GFMC results are taken from Ref. \cite{wiringa02}
after the subtraction of the Coulomb term of 0.7MeV\cite{pieper06}. The 
EIHH result\cite{orlandini06} doesn't contain the Coulomb interaction.
All the value are expressed in MeV.}
\begin{center}
\begin{tabular}{|c|c|c|}
\toprule
method & $E(^4{\rm He})$  & $E(^8{\rm He})$  \\
\hline
AFDMC & -27.13(10) & -23.6(5) \\
GFMC  & -26.93(1) & -23.6(1) \\
EIHH  & -26.85(2) & \\
\botrule
\end{tabular}
\label{tab:alpha}
\end{center}
\end{table}

Our trial function contains no tensor correlations and the variational
estimate is not even bound. 
The diffusion process enforced by the AFDMC method is capable of crossing 
the transition from an unbound to a bound system, leading to energy estimates which 
compare very well with the available GFMC results. Table \ref{tab:alpha} reports results for 
the alpha particle and the open shell nucleus $^8$He.
For the alpha particle AFDMC estimates are compared with GFMC and the Effective Interaction 
Hyperspherical Harmonic (EIHH) methods\cite{barnea00}. The AFDMC agreement with GFMC and EIHH for $^4$He is within 
about 1\% of the total energy. The agreement between AFDMC and GFMC for $^8$He is even better.

We have compared our results for $^{16}$O with other methods.
The variational FHNC/SOC\cite{fabrocini00},
and Cluster variational calculations\cite{pieper92} used the Argonne $v_{14}$
interaction. Our result for the energy, keeping just the first six operators,
is -90.8(1) MeV. The variational results keeping just those
same six operators are
-83.2 MeV from Variational Monte Carlo and -84.0 MeV from FHNC/SOC.
The AFDMC method seems to lower the energy by about 10\% with
respect to the two different variational results; however, the
variational wavefunctions were optimized with the full $v_{14}$ interaction
instead of our truncated interaction.


We then performed calculations for the $^{16}$O and for $^{40}$Ca with the Argonne $v_6'$ NN interaction. We 
chose this potential because it is a reprojected version of the more sophisticated Argonne $v_{18}$.
Its main deficiency is the lack of
a spin-orbit interaction and the three-body potential. 

\begin{table}
\caption{Computed Ground-State Energy in MeV of $^4$He, $^8$He, $^{16}$O 
and $^{40}$Ca for the Argonne $v_6'$ interaction. Experimental energies are also reported\cite{exp00}. 
We also calculated the energy of 28 nucleons in a periodic box to extrapolated the nuclear matter 
energy at equilibrium density as described in Ref. \cite{gandolfi07}
All the value are expressed in MeV.}
\begin{center}
\begin{tabular}{|c|c|c|c|c|}
\toprule
nucleus   & $E$       & $E/A$ & $E_{exp}$ & $E_{exp}/A$ \\
\hline
$^4$He    & -27.20(5) & -6.8  & -28.296   & -7.074 \\
$^8$He    & -23.6(5)  & -2.95 & -31.408   & -3.926 \\
$^{16}$O  & -100.7(4) & -6.29 & -127.619  & -7.98  \\
$^{40}$Ca & -272(2)   & -6.8  & -342.051  & -8.55 \\
nuclear matter &      & -12.8(1)   & & \\
\botrule
\end{tabular}
\label{tab:results}
\end{center}
\end{table}

Results are reported in table \ref{tab:results}, where it is reported also the energy of nuclear matter at 
the equilibrium density $\rho_0$=0.16fm$^{-3}$ calculated with
AFDMC\cite{gandolfi07}.

As expected the $v_6'$ interactions is not sufficient to build the total binding energy of $^{16}$O and of 
$^{40}$Ca.
This NN interaction gives
about 96\% of total binding energy for alpha particle, 75\% for $^8$He, 79\% for $^{16}$O 
and 79\% for $^{40}$Ca. 
Our $^{16}$O is unstable to break
up into 4 alpha particles, and our $^{40}$Ca has the same
energy of 10 alpha particles. This is behavior is consistent with the simple
pair counting argument of Ref. \cite{wiringa06}.
The surface energy coefficient in the Weizsacker formula, resulting from the comparison of the 
binding energies per nucleon of symmetrical nuclear matter and $^{40}$Ca is 20.5 MeV, not too far from 
the experimental value of 18.6 MeV.


We have extended the AFDMC method, used to perform calculations for neutron systems, to study 
finite nucleonic systems and have obtained agreement with
results from other methods for light nuclei.
The method easily accommodates
open-shell nuclei. We have calculated systems with up to
A=40 nucleons here. The computational time per imaginary time step scales 
as $A^3$. The total computational time depends on the desired quantity,
the distribution of excited states, and the quality and complexity
of the trial function just as in all other quantum Monte Carlo methods.
We believe that the results we have presented show that AFDMC in the fixed phase approximation 
has become a very powerful tool to solve large nuclear systems with realistic interactions. This opens up 
the possibility of calculating at an enriched accuracy heavy nuclei and asymmetric nuclear matter.

To be able to predict accurately the structure of nuclei, we must
use a more realistic Hamiltonian. The main two features missing from this
work are the three-body potential and spin orbit terms.

The Urbana-IX potential has been used by us in previous neutron
studies. This potential contains a 
spin-independent short range repulsion, and operator terms
based on the Fujita-Miyazawa model. The spin-independent part
and the so-called anticommutator terms lead
to terms with two- or fewer spin-isospin operators and can be included
immediately in AFDMC calculations.
The commutator terms have as well as the isospin exchange spin-orbit
require additional auxiliary fields to rewrite
them in terms of single-particle spin-isospin operators.
The other terms in a more realistic interaction are
included perturbatively in GFMC calculations. Initially, we will
also try to include them perturbatively. This will likely be
more difficult in AFDMC than in GFMC since our trial wave functions are
much simpler and much less accurate so that the perturbative estimates
will likely also be less accurate. The inclusion of these terms
is in progress and will be the subject of future works.

We thank G. Orlandini, W. Leidemann, and S.C. Pieper for providing us the
EIHH and GFMC results with $v_6'$ for comparison.
This work was supported in part by NSF grant PHY-0456609.
Calculations were performed on the
HPC facility "BEN" at ECT* in Trento under a grant for Supercomputing Projects.


\end{document}